\pdfoutput=1
\documentclass{llncs}
\usepackage{makeidx}  
\usepackage[utf8]{inputenc}
\usepackage{amssymb}
\usepackage{stmaryrd}
\usepackage{amsmath}
\usepackage{mathrsfs}
\usepackage{colonequals}
\usepackage{url}
\usepackage{bm}
\usepackage{cite}
\usepackage{xcolor}
\usepackage{chngcntr}
\usepackage{graphicx}
\usepackage{booktabs}

\renewcommand{\Pr}{\mathrm{Pr}}

\newcommand{\defeq}{\colonequals}

\newcommand{\cnf}{\mathrm{CNF}}

\newcommand{\kk}{\langle k \rangle}

\newcommand\pqe{\ensuremath{\mathop{\mathsf{PQE}}}}

\newcommand\AND{\mathsf{and}}
\newcommand\OR{\mathsf{or}}
\newcommand\NOT{\mathsf{not}}
\newcommand\lin{\mathsf{Lin}}

\newcommand\VARS{\mathsf{Vars}}

\newcommand\sat{\mathsf{SAT}}
\newcommand\dep{\mathsf{DEP}}

\newcommand\tgl{\mathsf{Toggle}}
\newcommand\nice{\mathsf{Nice}}

\newcommand\x{\mathrm{x}}

\newcommand{\NN}{\mathbb{N}}

\newcommand\false{\bot}

\newcommand\restr[2]{{%
  \kern-\nulldelimiterspace %
  #1 %
  _{|#2} %
  }}

\begin{document}

\mainmatter              

\title{Towards Deterministic Decomposable Circuits\\
for Safe Queries}

\author{Mikaël Monet\inst{1} \and
Dan Olteanu\inst{2}}

\institute{LTCI, Télécom ParisTech, Université Paris-Saclay, France,\\
Inria Paris; Paris, France \email{mikael.monet@telecom-paristech.fr}
\and
University of Oxford \email{dan.olteanu@cs.ox.ac.uk}}

\maketitle              

\begin{abstract}
There exist two approaches for exact probabilistic inference of UCQs on tuple-independent databases. 
In the \emph{extensional approach}, query evaluation is performed within a DBMS by exploiting the structure of the query.
In
the \emph{intensional approach}, one first builds a representation of the \emph{lineage} of the query on the database, then computes the probability of the lineage.
In this paper we propose a new technique to construct lineage representations as deterministic decomposable circuits in PTIME. 
The technique can apply to a class of UCQs that has been conjectured to separate the complexity of the two approaches.
We test our technique experimentally, and show that it succeeds on all the queries of this class up to a certain size parameter, i.e., over $20$ million queries. 

\end{abstract}

\section{Introduction}
\label{sec:introduction}
Probabilistic databases~\cite{suciu2011probabilistic} have been introduced in answer to the need to capture data uncertainty and reasoning about it. 
This uncertainty can come from various angles: imperfect sensor precision of scientific data, imprecise automatic processes (e.g., natural language processing, rule mining in knowledge bases), untrusted data sources (e.g., web crawling), etc. 
In their simplest and most common form, probabilistic databases
consist of a relational database where each tuple is annotated with a probability value that is supposed to represent how confident we are about having this tuple in the database. 
While a traditional (deterministic) database can only \emph{satisfy} or \emph{violate} a 
Boolean query, a probabilistic database has a certain probability of satisfying it.
Given a Boolean query $Q$ the \emph{probabilistic query evaluation problem for $Q$} ($\pqe(Q)$) then asks for the probability that the query holds on an input probabilistic database.
We measure the complexity of $\pqe(Q)$ as a function of the input database, hence considering that the Boolean query $Q$ is fixed.
This is known as \emph{data complexity}, and is motivated by the fact that the queries are usually much smaller than the data.

Unfortunately, even for very simple queries,  $\pqe(Q)$ can be intractable.
When $Q$ is a union of conjunctive queries (UCQ), a dichotomy result is provided by the work of Dalvi and Suciu~\cite{dalvi2012dichotomy}: either $Q$ is \emph{safe}
and $\pqe(Q)$ is PTIME, or $Q$ is not safe and $\pqe(Q)$ is \#P-hard. 
The algorithm to compute the probability of a safe UCQ exploits the first order structure of the query to find a so called \emph{safe query plan} (using extended relational operators that can manipulate probabilities) and can be implemented within a DBMS. 
This approach is referred to as \emph{extensional query evaluation}, or \emph{lifted inference}.

A second approach to $\pqe$ is \emph{intensional query evaluation} or \emph{grounded inference}, and consists of two steps. 
First, compute a representation of the \emph{lineage} of the
query $Q$ on the database $D$, which is a Boolean formula intuitively representing which tuples of~$D$ suffice to satisfy~$Q$.
Second, perform weighted model counting on the lineage to obtain the probability.
To ensure that model counting is tractable, we use the structure of the query to represent the lineage in tractable formalisms from the field of knowledge compilation, such as
read once Boolean formulas, free or ordered binary decision diagrams (OBDDs, FBDDs), deterministic decomposable normal forms (d-DNNFs), decision decomposable normal forms (dec-DNNFs), deterministic decomposable circuits (d-Ds), etc.
The main advantage of this approach compared to lifted inference is that the lineage can help explain the query answer. Moreover, having the lineage in a good knowledge compilation formalism can be useful for other applications: we could for instance change the tuples' probabilities and compute the new result easily, or compute the most probable state of the database that satisfies the query.

What we call the \emph{$q_9$ conjecture}, formulated by Dalvi, Jha, and Suciu~\cite{jha2013knowledge, dalvi2012dichotomy}, states that for safe queries, extensional query evaluation is strictly more powerful than the 
knowledge compilation approach.
Or in other words, that there exists a query which is safe (i.e., can be handled by the extensional approach) whose lineages on arbitrary databases cannot be computed in PTIME in a knowledge compilation formalism that allows tractable weighted model counting (i.e., cannot be handled by the intensional approach). 
Note that the conjecture depends on the tractable formalism that we consider.
The conjecture has recently been shown by Beame, Li, Roy, and Suciu~\cite{beame2017exact} to hold for the formalism of dec-DNNFs (including OBDDs and FBDDs), which captures the traces of modern model counting algorithms. 
Another independent result by Bova and Szeider~\cite{bova2017circuit} shows that the conjecture also holds when we consider the class of \emph{deterministic structured negation normal forms} (d-SDNNFs), which are d-DNNFs that follow the structure of a \emph{v-tree}~\cite{pipatsrisawat2008new}.
However the question is still open for more expressive formalisms, namely, d-DNNFs and d-Ds. 
Maybe the conjecture fails for such expressive formalisms, i.e., maybe the reason why $\pqe$ is PTIME for safe queries is because we can build deterministic decomposable circuits in PTIME for them?

In this paper we focus on a class of queries (the $\mathcal{H}$-queries) that was conjectured in~\cite{jha2013knowledge, dalvi2012dichotomy} to separate the two approaches and that was used to prove the conjecture for
dec-DNNFs~\cite{beame2017exact} and d-SDNNFs~\cite{bova2017circuit}. 
Our first contribution is to develop a new technique to build d-DNNFs and d-Ds in polynomial time for the $\mathcal{H}$-queries, based on what we call \emph{nice Boolean functions}. 
Because we were not able to prove that this technique works for all
the safe $\mathcal{H}$-queries, our second contribution is to test this
technique with the help of the SAT solver Glucose~\cite{audemard2009predicting} on all the $\mathcal{H}$ queries up to a certain size parameter, that we generated automatically.
We found no query on which it does not work. 
Interestingly, we found a few queries for which we can build d-Ds with a single internal negation at the very top, whereas we do not know if we can build d-DNNFs (could these queries
separate UCQ(d-DNNF) and UCQ(d-D)?).
We conjecture that this technique can build d-Ds for all safe $\mathcal{H}$-queries.

To do this analysis, we had to solve a task of independent interest, namely, computing explicitly the list of all inequivalent monotone Boolean functions on $7$ variables.
This task had previously been undertaken by Cazé, Humphries, and Gutkin~\cite{caze2013dendrites} and by Stephen and Yusun~\cite{stephen2014counting}.
We reused parts of the code from~\cite{caze2013dendrites} and confirmed the number of such functions: $490,013,148$.

\paragraph*{Paper structure}
We start our presentation with preliminaries in Section~\ref{sec:preliminaries}. We then define the $\mathcal{H}$-queries in Section~\ref{sec:hk} and review what is known about them.
In Section~\ref{sec:nice} we introduce our technique, and we experimentally demonstrate its effectiveness in Section~\ref{sec:experiments}.
Our code and all the functions are available online~\cite{monet2018code}.

\section{Preliminaries}
\label{sec:preliminaries}
We will consider in this work the most commonly used model for probabilistic databases: the \emph{tuple-independent} model, where each tuple is annotated with a probability
of being present or absent, assuming independence across tuples:

\begin{definition}
A \emph{tuple-independent (TID) database} is a pair $(D,\pi)$ consisting of a relational instance $D$ and a function $\pi$
mapping each tuple $t \in D$ to a rational probability $\pi(t) \in [0;1]$. A TID instance
$(D, \pi)$ defines a probability distribution $\Pr$ on $D' \subseteq D$, where
$\Pr(D') \defeq \prod_{t \in D'} \pi(t) \times \prod_{t \in D \backslash D'} (1 -
  \pi(t))$.
Given a Boolean query $Q$, the \emph{probabilistic query evaluation problem for $Q$} ($\pqe(Q)$) asks, given as input a TID instance $(D, \pi)$, the probability that $Q$ is
  satisfied in the distribution $\Pr$. That is, formally, $\Pr(Q, (D, \pi)) \defeq \sum_{D' \subseteq D\text{~s.t.~}D' \models Q} \Pr(D')$. 
\end{definition}

Dalvi and Suciu~\cite{dalvi2012dichotomy} have shown a dichotomy result on UCQs for $\pqe$: either $Q$ is \emph{safe}
and $\pqe(Q)$ is PTIME, or $Q$ is not safe and $\pqe(Q)$ is \#P-hard. 
Moreover they show that all the safe queries can be handled by the \emph{extensional approach}, i.e., by using the structure of the query to compute the probability.
Due to space constraints, we point to~\cite{dalvi2012dichotomy} for a presentation of their algorithm to compute the probability of a safe query, though it is not strictly necessary to understand the current paper.
We denote by $\text{UCQ(P)}$ the set of safe UCQs (hence which corresponds to the set of tractable UCQs if P $\neq$ \#P).

By contrast, in the \emph{intentional approach}, one first computes a representation of the \emph{lineage} $\lin(Q,D)$ of the
query $Q$ on the instance $D$:

\begin{definition}
\label{def:lineage}
The \emph{lineage} of a Boolean query $Q$ over~$D$ is a Boolean
formula $\lin(Q, D)$ on the tuples of~$D$
mapping each Boolean valuation $\nu: D \to \{0, 1\}$
to~$1$ or~$0$ depending on whether $D_\nu$ satisfies~$Q$ or not, where
$D_\nu \colonequals \{t \in D \mid \nu(t) = 1\}$.
\end{definition}

The lineage can be represented with any formalism that represents Boolean functions (Boolean formulas, BDDs, Boolean circuits, etc), but the crucial idea is
to use a formalism that allows tractable probability computation.
In this work we will specifically focus on \emph{deterministic decomposable circuits} (d-Ds) and 
\emph{deterministic decomposable normal forms}~\cite{darwiche2001tractability} (d-DNNFs).

\begin{definition}
Let $C$ be a Boolean circuit (featuring $\AND$, $\OR$, $\NOT$, and variable gates). 
An $\AND$-gate $g$ of $C$ is \emph{decomposable} if for every two input gates
$g_1\neq g_2$ of $g$ we have $\VARS(g_1) \cap \VARS(g_2) = \emptyset$, where $\VARS(g)$ denotes the set of variable gates that have a directed path to~$g$ in~$C$.
We call $C$ \emph{decomposable} if each $\AND$-gate is.
An $\OR$-gate $g$ of $C$ is \emph{deterministic} if there
is no pair $g_1\neq g_2$ of input gates of~$g$ and valuation $\nu$ of the variables such
that 
$g_1$ and $g_2$ both evaluate to~$1$ under~$\nu$.
We call $C$ \emph{deterministic} if each $\OR$-gate is.
A \emph{negation normal form} (NNF) is a circuit in which the inputs of $\NOT$-gates are always variable gates.
\end{definition}

Probability computation is in linear time for d-Ds (hence, for d-DNNFs):
to compute the probability of a d-D, compute by a bottom-up pass the probability of each gate,
where $\AND$ gates are evaluated using $\times$, $\OR$ gates using $+$, and $\NOT$ gates using $1-x$.
While there does not seem to be any interest in using d-DNNFs rather that d-Ds for probabilistic databases, we are also interested by d-DNNFs from a knowledge compilation point of view, as it is currently not known if \mbox{d-Ds} are strictly more succinct than d-DNNFs. 
We write UCQ(d-DNNF) (resp., UCQ(d-D)) to denote the set of UCQs $Q$ such that for any
database instance~$D$, we can compute in polynomial time (in data complexity) a d-DNNF (resp., \mbox{d-D}) representation of $\lin(Q,D)$. 
For a study of the intensional approach using weaker
formalisms for Boolean functions (read once formulas, ordered and free binary decision diagrams), see~\cite{jha2013knowledge}.
Hence we have:
\begin{equation}
\text{UCQ(d-DNNF)} \subseteq \text{UCQ(d-D)} \subseteq \text{UCQ(P)}
\end{equation}

Dalvi, Jha, and Suciu~\cite{jha2013knowledge,dalvi2012dichotomy} conjectured that the inclusion $\text{UCQ(d-D)} \subseteq \text{UCQ(P)}$ is strict, i.e., that the extensional approach is strictly more powerful than the intensional approach, and proposed a candidate query to separate these classes (named $q_9$ and that we define in the next section).
The purpose of this paper is to study this conjecture.

\section{The $\mathcal{H}$-queries}
\label{sec:hk}
We define in this section the $\mathcal{H}$-queries and review what is known about them. The building blocks of these queries are the queries $h_{ki}$, which were first defined in the work of Dalvi and Suciu to show the hardness of UCQs that are not safe:

\begin{definition}
\label{def:hk}
Let $k \in \NN$, $k \geq 1$. The queries $h_{ki}$ for $0 \leq i \leq k$ are defined by:
\begin{itemize}
	\item $h_{k0} = \exists x \exists y~R(x) \land S_1(x,y)$;
	\item $h_{ki} = \exists x \exists y~S_i(x,y) \land S_{i+1}(x,y)$ for $1 \leq i < k$;
	\item $h_{kk} = \exists x \exists y~S_k(x,y) \land T(y)$.
\end{itemize}
\end{definition}

We define the $\mathcal{H}$-queries to be combinations of queries $h_{ki}$, as in~\cite{beame2017exact}:

\begin{definition}
	For $k \geq 1$, we define the set of variables $\kk \defeq \{0,\ldots,k\}$.
	Given a Boolean function $\phi$ on variables
	$\kk$, we define the Boolean query $Q_k^\phi$ to be the query represented by the first order formula
	$\phi[0 \mapsto h_{k0}, \ldots, k \mapsto h_{kk}]$, i.e., $\phi$ where we substituted each variable $i \in \kk$ by the formula $h_{ki}$.
\end{definition}

The query class $\mathcal{H}_k$ (resp., $\mathcal{H}^+_k$) is then the set of queries $Q^\phi_k$ when $\phi$ ranges over all 
Boolean functions (resp., monotone Boolean functions) on variables $\kk$.
We finally define $\mathcal{H}$ (resp., $\mathcal{H}^+$) to be $\bigcup\limits_{k=1}^\infty \mathcal{H}_k$ (resp., $\bigcup\limits_{k=1}^\infty \mathcal{H}^+_k$).
Observe that the queries in $\mathcal{H}^+$ are in particular UCQs.

\begin{example}
	\label{expl:q9}
	Let $k=3$, and $\phi_9$ be the monotone Boolean function $(2 \vee 3) \wedge (0 \vee 3) \wedge (1 \vee 3) \wedge (0 \vee 1 \vee 2)$. Then $Q_3^{\phi_9}$ 
	represents the query $q_9 = (h_{32} \vee h_{33}) \wedge (h_{30} \vee h_{33}) \wedge (h_{31} \vee h_{33}) \wedge (h_{30} \vee h_{31} \vee h_{32}) \in \mathcal{H}^+_3$, 
	which is safe and was conjectured
	in~\cite{jha2013knowledge, dalvi2012dichotomy} not to be in UCQ(d-D).
\end{example}

To study the $\mathcal{H}$-queries, we need the following notions on Boolean functions:

\begin{definition}
Let $\phi$ be a Boolean function on variables $\kk$.
We will always consider a valuation $\nu$ of $\kk$ simply as the set of variables that $\nu$ maps to $1$.
We write $\sat(\phi)$ the set of satisfying valuations of $\phi$.
We say that $\phi$ \emph{depends on variable $l \in \kk$} if there exists a valuation $\nu \subseteq \kk$
such that $\phi (\nu \cup \{l\}) \neq \phi (\nu \setminus \{l\})$.
We write $\dep(\phi) \subseteq \kk$ for the set of variables on which $\phi$ depends.
We call $\phi$ and $Q^\phi_k$ \emph{nondegenerate} if $\dep(\phi) = \kk$ (and \emph{degenerate} otherwise).
\end{definition}

Then, if $\phi$ is degenerate (i.e., does not depend on all its $k+1$ variables), $Q_k^\phi$ is safe and is in UCQ(d-DNNF) (in fact, even in UCQ(OBDD)):

\begin{proposition}[Theorem 3.12 of~\cite{beame2017exact}, or Lemma 3.8 of~\cite{fink2016dichotomies}]
\label{prp:strict_subset}
	Let $k \geq 1$, and $Q^\phi_k \in \mathcal{H}_k$ with $\dep(\phi) \subsetneq \kk$. Then
	$Q_k^\phi \in$ UCQ(d-DNNF).
\end{proposition}

This is in contrast to when $\phi$ is nondegenerate. Indeed,
Beame, Li, Roy, and Suciu then show~\cite{beame2017exact} that $Q^\phi_k$ do not admit polynomial sized decision decomposable NNFs (dec-DNNF). A dec-DNNF is a d-DNNF in which the determinism of $\OR$ gates
is restricted to simply choosing the value of a variable~\cite{huang2005dpll, huang2007language}. That is, each $\OR$ gate is of the form $(v \land g) \lor (\lnot v \land g')$ for some variable $v$. 
In fact, they show a lower bound for more general representations than dec-DNNFs, namely, for what they called \emph{Decomposable Logic Decision Diagrams} (DLDDs), which generalise dec-DNNFs
in that they allow negations at arbitrary places and also allow decomposable binary operator gates. 
When $\phi$ is monotone, another independent lower bound by Bova and Szeider~\cite{bova2017circuit} tells us than we cannot impose \emph{structuredness} either (i.e., use d-SDNNFs) when $\phi$ is nondegenerate.
These results mean that for such queries, one cannot restrict too much the expressivity of determinism.
The question is then: do the nondegenerate queries have polynomial sized d-DNNFs (or d-Ds)?

Let us first see what the dichotomy theorem tells us about $\mathcal{H}$-queries that are nondegenerate. 
We shall restrict our attention to monotone functions now, i.e., to queries in $\mathcal{H}^+$, because the dichotomy theorem applies only to UCQs, and $Q^\phi_k$ is not a UCQ
when $\phi$ is not monotone.
We need to define the \emph{CNF lattice} of $\phi$:

\begin{definition}
\label{def:lattices}
	Let $\phi$ be a monotone Boolean function on variables $\kk$ such that $\dep(\phi)=\kk$, and let $F_\cnf = C_0 \land \ldots \land C_n$ 
	 be the 
	(unique) minimized CNF representing $\phi$, where we see each clause simply as the set of variables that it contains. 
	For $\mathbf{s} \subseteq \langle n \rangle$, we define $d_\mathbf{s} \defeq \bigcup\limits_{i \in \mathbf{s}} C_i$. 
	Note that $d_\emptyset$ is $\emptyset$, and that we can have $d_\mathbf{s} = d_\mathbf{s'}$ for
	$\mathbf{s} \neq \mathbf{s'}$.
	The \emph{CNF lattice of $\phi$} is the lattice $(L,\leq)$, where 
	$L$ is $\{ d_\mathbf{s} \mid \mathbf{s} \subseteq \langle n \rangle \}$, and where $\leq$ is reversed set inclusion.
	In particular, the top element $\hat{1}$ of $L_\cnf$ is $\emptyset$, while its bottom element $\hat{0}$ is $\kk$ (because $\phi$ depends on all
	the variables, hence each variable is in at least one clause).
\end{definition}

\begin{figure}
\centering
\includegraphics[scale=0.6]{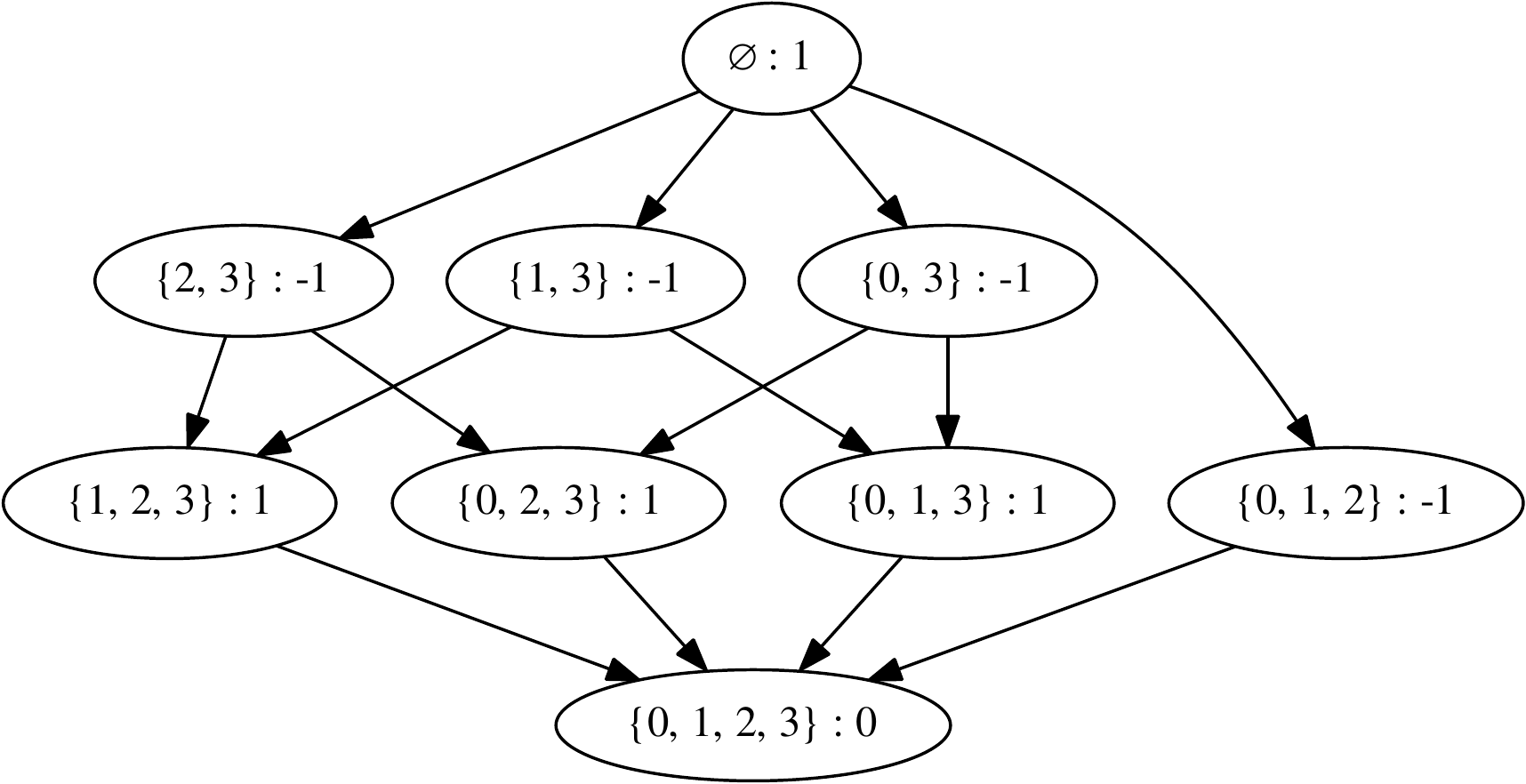}
\caption{CNF lattice of $\phi_9$ with the values $\mu(n,\hat{1})$ for each node $n$.}
\label{fig:lattice}
\end{figure}

\begin{example}
\label{expl:lattice}
	The Hasse diagram of the CNF lattice of $\phi_9$ is shown in Figure~\ref{fig:lattice} (ignore for now the values at the right inside the nodes).
\end{example}

The criterion of Dalvi and Suciu is based on the \emph{Mobius function}~\cite{stanley2011enumerative} of the CNF lattice $L$ of $\phi$.
The Mobius function $\mu: L \times L \to \mathbb{Z}$ on $L$ is defined on pairs $(u,v)$ with $u \leq v$ by $\mu(u,u) = 1$ and 
$\mu(u,v) = - \sum\limits_{u < w \leq v} \mu(w,v)$ for $u < v$.
The dichotomy theorem for UCQs then states:
\begin{enumerate}
	\item If $\mu_L(\hat{0},\hat{1}) \neq 0$, then $Q_k^\phi$ is \#P-hard. Hence, if P is different from \#P, we have $Q_k^\phi \notin$ UCQ(d-D).

	\item If $\mu_L(\hat{0},\hat{1}) = 0$, then $Q_k^\phi$ is PTIME. 
		We do not know if $Q_k^\phi \in \text{UCQ(d-DNNF)}$ or $Q_k^\phi \in \text{UCQ(d-D)}$.
		But by what precedes, we know that $Q_k^\phi \notin \text{UCQ(dec-DNNF)}$ and $Q^\phi_k \notin \text{UCQ(d-SDNNF)}$.
	\end{enumerate}

Our goal is to investigate, when we are in the second case ($\phi$ is monotone, nondegenerate and safe), for which functions $\phi$ we can build d-DNNFs or d-Ds.

\section{Nice Boolean Functions}
\label{sec:nice}
In this section we present a technique to prove that some queries $Q_k^\phi \in \mathcal{H}_k$ are in $\text{UCQ(d-DNNF)}$. 
This will in particular apply to the query $q_9$ (which, we recall, was conjectured not to be in UCQ(d-D)).
Our goal is to rewrite $\phi$ as $\phi_0 \lor \ldots \lor \phi_k$, where the $\phi_i$ are mutually exclusive ($\phi_i \land \phi_j \equiv \false$ for $i\neq j$) 
and depend on strict subsets $S_i \subsetneq \kk$ of $\kk$. When such a rewriting exists, we say that $\phi$ is \emph{nice}:

\begin{definition}
	\label{def:nice}
	Let $\phi$ be a Boolean function on variables $\kk$. We call $\phi$ \emph{nice} if there exist 
	strict subsets $S_i$ of $\kk$ and mutually exclusive Boolean functions (not necessarily monotone) $\phi_i$ for $0 \leq i \leq k$
	such that $\dep(\phi_i) = S_i$ and $\phi \equiv \bigvee\limits_{i=0}^k \phi_i$.
\end{definition}

Observe that allowing to have more (or less) than $k+1$ functions $\phi_i$ would not change the definition of being nice.
Also, note that if $\phi$ is degenerate, then $\phi$ is trivially nice.

\begin{example}
	\label{expl:q9-nice}
	The function $\phi_9$ is equivalent to the mutually exclusive disjunction $\phi_0 \vee \phi_1 \vee \phi_2 \vee \phi_3$, where
		$\phi_0 \equiv 0 \land \lnot 2 \land 3$; $\phi_1 \equiv \lnot 1 \land 2 \land 3$;  $\phi_2 \equiv \lnot 0 \land 1 \land 3$; and
		 $\phi_3 \equiv 0 \land 1 \land 2$.
	Moreover for $0 \leq i \leq 3$ we have $\dep(\phi_i) \subsetneq \langle 3 \rangle$, hence $\phi_9$ is nice.
\end{example}

When $\phi$ is nice, we can express $Q_k^\phi$ as $\bigvee\limits_{i=0}^{k} Q_k^{\phi_i}$, thus showing that $Q_k^\phi \in \text{UCQ(d-DNNF)}$. 
Indeed, given a database $D$, we can use Proposition~\ref{prp:strict_subset} to construct in PTIME a 
d-DNNF $C^{\phi_i}$ representing $\lin(Q_k^{\phi_i},D)$ for  each $i \in \{0,\ldots, k\}$,
and then build the d-DNNF $C^\phi = \bigvee\limits_{i=0}^{k} C^{\phi_i}$ that represents $\lin(Q_k^\phi,D)$.
In other words, the following holds:

\begin{proposition}
Let $k \in \NN$ and $\phi$ be a Boolean function on variables $\kk$.
If $\phi$ is nice, then $Q_k^\phi \in \text{UCQ(d-DNNF)}$.
\end{proposition}

Hence $q_9 \in $UCQ(d-DNNF).
This result shows that, for all queries $Q^\phi_k$ where $\phi$ is nice, we can compute a d-DNNF representation of their lineage in PTIME, and hence compute their probability efficiently.
We do not know to which queries this technique can be applied. Moreover also we have the following corollary:
\begin{corollary}
	Let $k \in \NN$ and $\phi$ be a Boolean function on variables $\kk$.
	If $\lnot \phi$ is nice, then $Q_k^\phi \in$ UCQ(d-D).
\end{corollary}

We will call \emph{co-nice} a function $\phi$ such that $\lnot \phi$ is nice.

\section{Experiments}
\label{sec:experiments}
We have presented a technique
that can be used to show that some queries are in UCQ(d-DNNF) or
UCQ(d-D), but we have not characterized the queries to which it
applies. In this section, we present our experiments that
show that for all $k \in \{1,\ldots,6\}$, every nondegenerate monotone function $\phi$ for which $Q^\phi_k$ is safe is either nice or co-nice.
Hence all the safe queries in $\mathcal{H}^+_k$ for $k \in \{1,\ldots,6\}$ are in UCQ(d-D).
This suggests that UCQ(d-D) $=$ UCQ(P), or at least that any counterexample query in $\mathcal{H}^+$ must be in $\mathcal{H}^+_k$ for $k \geq 7$.

We used a machine with $40$ x$86$ $64$ CPUs of 2.6 GHz and $512$~GB RAM. 
The code was written in Python~2.7.12 and parallelized using Python's multiprocessing library.
We explain briefly how we generated all the functions in $\mathcal{H}^+_k$ for $k \in \{1,\ldots,6\}$ in Section~\ref{sec:generate}, then explain how
we tested niceness of these functions in Section~\ref{sec:test_niceness}.

\subsection{Generating $\mathcal{H}^+_k$}
\label{sec:generate}

We started by generating the set $R(k)$ of all monotone Boolean functions on variables $\kk$ \emph{up to isomorphism}, that is, up to renaming the variables.
The size of $R(k)$ corresponds to the OEIS sequence A003182, which is only known up to $k=6$ 
(computed in~\cite{caze2013dendrites} and in~\cite{stephen2014counting}). 
We used parts of the code from~\cite{caze2013dendrites}
to generate all functions in $R(k)$ for $k$ in $\{1,\ldots,6\}$.
We then filtered $R(k)$ to obtain the set of functions that are nondegenerate. 
Then we tested whether $Q^\phi_k$ is safe by computing the CNF lattice of $\phi$ and checking that $\mu(\hat{0}, \hat{1}) = 0$. 
Let us call $\mathit{SND}(k)$ the set of remaining functions (that is, the functions that are safe and nondegenerate). 
It took about $2$ weeks (using the $40$ CPUs) to compute the explicit lists of all the functions in $R(k)$ and $\mathit{SND}(k)$ for $k \in \{1,\ldots,6\}$,
and the sizes of these sets can be found in Table~\ref{tab:results}.
We next explain how we tested the niceness of each function in $\mathit{SND}(k)$.

\subsection{Testing Niceness}
\label{sec:test_niceness}

Let us call \emph{boxes} the functions $\phi_i$ used in Definition~\ref{def:nice}.
That is, $\phi$ is nice if and only if we can partition its satisfying valuations into $k+1$ ordered boxes (we allow some boxes to be empty), where the $i$-th box has a \emph{symmetry around variable $i$}:

\begin{definition}
	Let $\nu \subseteq \kk$ be a valuation of $\kk$, and $l \in \kk$ be a variable. 
	We define the valuation $\tgl(\nu,l)$ to be the valuation $\nu \cup \{l\}$ if $l \notin \nu$ and
	$\nu \setminus \{l\}$ if $l \in \nu$.
	We say that a set $B$ of valuations of $\kk$ \emph{has a symmetry around variable $l$} if for every valuation $\nu \subseteq \kk$ we have $\nu \in B$ iff $\tgl(\nu,l) \in B$.
\end{definition}

To check if $\phi$ is nice, we build a CNF $\nice(\phi)$ that expresses exactly that $\sat(\phi)$ can be partitioned nicely, 
i.e., $\nice(\phi)$ is satisfiable if and only if $\phi$ is nice. We can then use a SAT solver.

\begin{definition}
	Let $k \geq 1$ and $\phi$ be a Boolean function on $\kk$. We define the CNF $\nice(\phi)$ as follows. 
	Its set of variables is $\{\x_\nu^l \mid \nu \in \sat(\phi) \text{ and } l \in \kk \}$, where $\x_\nu^l$ intuitively expresses that $\nu$ is put in box $l$. Its set of clauses is:
	\begin{enumerate}
		\item For each $\nu \in \sat(\phi)$, the clause $\bigvee\limits_{l=0}^{k} \x_\nu^l$, expressing the valuation $\nu$ must be put in at least one box;
		\item For each $\nu \in \sat(\phi)$ and $l,l' \in \{0,\ldots,k\}$ with $l \neq l'$, the clause $\lnot \x_\nu^l \lor \lnot \x_\nu^{l'}$,
			expressing that the valuation $\nu$ is in at most one box;
		\item For each $\nu \in \sat(\phi)$ and $l \in \{0,\ldots,k\}$, then:
			\begin{enumerate}
				\item If $\tgl(\nu,l) \notin \sat(\phi)$, the clause $\lnot \x_\nu^l$;
				\item Else, the clause $\lnot \x_\nu^l \lor \x_{\tgl(\nu,l)}^l$.
			\end{enumerate}
			This ensures that the box $l$ has a symmetry around $l$.

	\end{enumerate}
\end{definition}

\begin{proposition}
	\label{prp:nice_is_nice}
	$\phi$ is nice iff $\nice(\phi)$ is satisfiable.
\end{proposition}

Now for each function $\phi$ in $\mathit{SND}(k)$, we constructed the CNF formula $\nice(\phi)$,
and used the SAT solver Glucose~\cite{audemard2009predicting} to determine if it is satisfiable. 
If $\nice(\phi)$ is satisfiable then $Q^\phi_k \in \text{UCQ(d-DNNF)}$ and we store $\phi$ in $N(k)$. 
If it is not we give the formula $\nice(\lnot \phi)$ to Glucose. 
If this formula is satisfiable then $\phi$ is co-nice and $Q^\phi_k \in \text{UCQ(d-D)}$ (but we do not know if it is in UCQ(d-DNNF)) and we store $\phi$ in $\text{\emph{co-N}}(k)$. 
If $\nice(\lnot \phi)$ is not satisfiable then $\phi$ is in $\mathit{BAD}(k)$ and we
do not know if $Q^\phi_k$ is in UCQ(d-D).
The results of these experiments are displayed in Table~\ref{tab:results}, and, as we found no function in $\mathit{BAD}(k)$, imply:

\setlength{\tabcolsep}{10pt}
\begin{table}[t]
	\caption{Results of our experiments. The meaning of columns is explained after Proposition~\ref{prp:nice_is_nice}.}

  \centering
\begin{tabular}{crrrrr}
    \toprule
    $k$ &  $|R(k)|$ & $|\mathit{SND}(k)|$ & $|N(k)|$ & $|\text{\emph{co-N}}(k)|$ & $|\mathit{BAD}(k)|$ \\
    \midrule
    $1$ &  $5$ & $0$ & $0$ & $0$ & $0$ \\
    $2$ &  $10$ & $0$ & $0$ & $0$ & $0$ \\
    $3$ &  $30$ & $2$ & $2$ & $0$ & $0$ \\
    $4$ &  $210$ & $25$ & $25$ & $0$ & $0$ \\
    $5$ &  $16,353$ & $2,531$ & $2,529$ & $2$ & $0$ \\
    $6$ &  $490,013,148$ & $21,987,161$ & $21,987,094$ & $67$ & $0$ \\
    \bottomrule
\end{tabular}
\label{tab:results}
\end{table}

\begin{proposition}
	All the safe queries in $\mathcal{H}^+_k$ for $k \in \{1,\ldots,6\}$ are in UCQ(d-D).
\end{proposition}

We give here one of the $2$ functions that are in $\text{\emph{co-N}}(5)$, $\phi_{\mathrm{co-N1}} \defeq 
24 \land 034 \land 013 \land 12 \land 15 \land 05 \land 35 \land 23 \land 02 \land 25 \land 014 \land 45$
 where we write, for instance, $014$ to mean $0 \lor 1 \lor 4$. 
 Could $\phi_{\mathrm{co-N1}}$ separate UCQ(d-DNNF) from UCQ(d-D)?

\section{Conclusion}
\label{sec:conclusion}
We have introduced a new technique to construct deterministic decomposable circuits for safe $\mathcal{H}$-queries and have experimentally demonstrated its effectiveness 
on the first $20$ million such queries.
We conjecture that this technique can build d-Ds for all safe $\mathcal{H}$-queries. We leave open many intriguing questions:
\begin{itemize}
	\item For the $\mathcal{H}$-queries, can we use the DNF lattice instead of the CNF lattice to decide if the query is safe~\cite{cstheory}?
	\item Can we show (unconditionally to P $\neq$ \#P) that if $Q^\phi_k$ is not safe then $\phi$ is not nice?
	\item What is the link between our technique and the notion of \emph{d-safety} defined in~\cite{jha2013knowledge}?
	\item Do the queries in $\text{\emph{co-N}}$ separate UCQ(d-DNNF) from UCQ(d-D)?
\end{itemize}

We have put online our code, and the complete list of all the functions studied in Section~\ref{sec:test_niceness}, which we believe
could be useful for people studying the $\mathcal{H}$-queries. 
It can be used, for instance, to enter by hand a monotone Boolean function $\phi$ and check if $Q^\phi_k$ is safe of not, draw its CNF lattice, check if $\phi$ is nice, etc.

\paragraph*{Acknowledgements.}
We are grateful to Antoine Amarilli for careful proofreading of the article (and for the many discussions about the code),
 to Romain Cazé for pointing out to us his code to generate monotone Boolean functions,
and to Stephen Tamon for referring us to Romain Cazé.
The second author would like to acknowledge Guy van den Broeck for initial discussions on the q9 conjecture. The fact that q9 is expressible as a succinct d-D has been already known to him and has been mentioned in several of his presentations prior to this article.
This work was partly funded by the Télécom ParisTech Research Chair on Big Data and Market Insights,
 and by the EPSRC platform grant DBOnto (L012138) that funded Mikaël's research visit at Oxford.

\bibliographystyle{splncs}
\bibliography{main}

\end{document}